\documentclass[11pt]{article}
\usepackage{graphics}
\usepackage{psfrag}
\usepackage{epsfig}
\usepackage{epsf}
\usepackage{float}
\usepackage{amssymb,stmaryrd,latexsym}
\newcommand{\boldpi}{\mbox{\boldmath $\pi$}}
\newcommand{\boldtau}{\mbox{\boldmath $\tau$}}
\newcommand{\boldT}{\mbox{\boldmath $T$}}

\newcommand{\beq}{\begin{equation}}

\newcommand{\eeq}{\end{equation}}
\newcommand{\beqa}{\begin{eqnarray}}
\newcommand{\eeqa}{\end{eqnarray}}

\newcommand{\barr}[1]{\not\mathrel #1}
\newcommand{\vs}{\vspace{-0.2cm}}

\begin{document}

\hfill {\small FZJ--IKP(TH)--2005--31, HISKP-TH-05/23}

\vspace{1in}

\begin{center}

{\Large\bf Towards a field theoretic understanding of {\boldmath$NN\to NN\pi$}}

\end{center}

\vspace{.3in}

\begin{center}

{\large V. Lensky$^{1,2}$, V. Baru$^{2}$, J. Haidenbauer$^1$,  C.~Hanhart$^1$,\\ 
A.~E. Kudryavtsev$^2$, and U.-G. Mei\ss ner$^{1,3}$
}

\bigskip

 $^1$Institut f\"{u}r Kernphysik, Forschungszentrum J\"{u}lich GmbH,\\ 
 D--52425 J\"{u}lich, Germany \\

\bigskip 

 $^2$Institute of Theoretical and Experimental Physics,\\
 117259, B. Cheremushkinskaya 25, Moscow, Russia\\

\bigskip 

 $^3$Helmholtz-Institut f\"{u}r Strahlen- und Kernphysik (Theorie), 
Universit\"at Bonn\\ 
 Nu{\ss}allee 14-16, D--53115 Bonn, Germany 
\end{center}

\vspace{.6in}

\thispagestyle{empty} 

\begin{abstract}
\noindent 
We study the production amplitude for the reaction $NN\to NN\pi$ up to
next--to--leading  order in chiral perturbation theory using a counting scheme
that takes into account the large scale introduced by the initial momentum. In
particular we investigate a subtlety that arises once the leading loop
contributions are convoluted with the $NN$ wavefunctions as demanded by the
non--perturbative nature of the $NN$ interaction.
We show how to properly identify the irreducible contribution of loop diagrams
in such type of reaction.
The net effect of the inclusion of all next-to-leading order 
loops is to enhance the leading
rescattering amplitude by a factor of 4/3, bringing its contribution to the
cross section for $pp\to d\pi^+$ close to the experimental value.  
\end{abstract}

\vfill

\pagebreak


\section{Introduction}

The highly accurate data for pion production in nucleon--nucleon collisions
close to the production threshold are a challenge for theoreticians.
When the first close to threshold data for the total cross section
of the reaction $pp \to pp\pi^0$ appeared in 1990, existing models fell short by
a factor of 5--10 \cite{kur,MuS}. Many different mechanisms were proposed to cure this
discrepancy: heavy meson exchanges 
\cite{LuR}, (off-shell) pion rescattering 
\cite{HO,Han1}, 
excitations of baryon resonances \cite{Pena}, and  pion emission from
exchanged mesons \cite{CP2b}. The total cross sections
for the reactions $pp\to pn\pi^+$ and $pp\to d\pi^+$
on the other hand could always be described within a factor of 2 
 ---  the amplitude is dominated by the isovector rescattering contribution \cite{kur}.
 For a recent review see Ref. \cite{report}.
 
 To resolve the situation various groups started to investigate $NN\to NN\pi$
 using chiral perturbation theory (ChPT). As an effective field theory (EFT) it is
 to be free of any ambiguities and people expected that now the relevant
 physics of $NN\to NN\pi$ could be identified.  As a big surprise to many,
 however, it turned out that, when naively using the original power counting
 by Weinberg \cite{swein1}, the discrepancy between
 theory and data became even larger at next--to--leading order (NLO) for
 $pp\to pp\pi^0$ \cite{park} as well as for $pp\to d\pi^+$
 \cite{unserd}. At the same time it was already realized that a
 modified power counting is nescessary to properly take care of the large
 momentum transfer characteristic for pion production in $NN$ collisions,
 however, when applied in its original formulation the results were basically
 the same for neutral \cite{bira1} as well as charged pions \cite{rocha}.
  Even worse, the corrections at one--loop order
 (next--to--next--to--leading order (N$^2$LO) in the standard counting) turned
 out to be even larger than the NLO corrections, indicating a divergence of
 the chiral expansion \cite{dmitrasinovic,ando}.

 Recently there were two developments: one that focused on some technical
 aspects related to the evaluation of the matrix elements \cite{toy,toy1} and
 another regarding the power counting for the large momentum transfer
 reactions.  Formal inconsistencies of the naive power counting using the
 heavy baryon scheme were pointed out in Ref.~\cite{BKMprod}.
In addition the ideas formulated in Refs.
 \cite{bira1,rocha} were further developed and improved --- it was especially
 recognized how to properly estimate loop contributions. This scheme
was implemented in Refs.  \cite{pionprod,norbertandme}
 --- the essential features are described in detail below. The basic
 conclusion was that an ordering scheme exists for the reactions $NN\to NN\pi$
 that can lead to a convergent series. However, so far full calculations
 (including the distortions due to the $NN$ interactions) within this scheme
 exist only for the production of $p$--wave pions \cite{pionprod}.

\begin{figure}[t]
\begin{center}
\epsfig{file=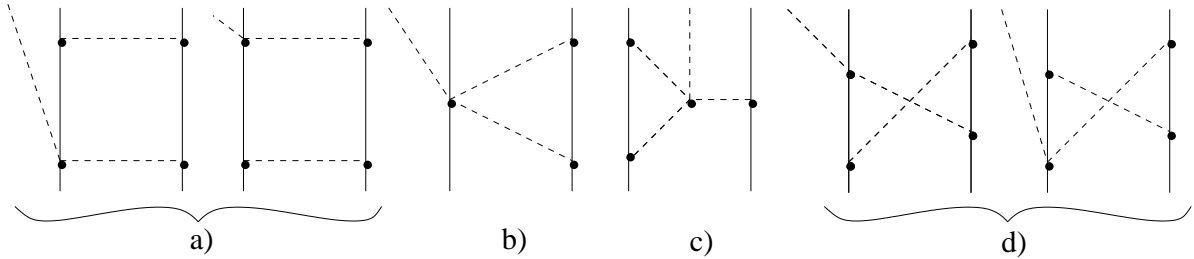, height=3.5cm, angle=0}
\caption{Leading loop diagrams for $NN\to NN\pi$. Here dashed lines denote
  pions and solid lines denote nucleons.}
\label{diagram}
\end{center}
\end{figure}

In Ref. \cite{norbertandme} it was demonstrated by explicit evaluation of the
leading loop contributions (shown in figure \ref{diagram}(b)--(d)) how the
presence of the large momentum scale influences loops. The central findings of
that work were that it is possible to define an ordering scheme for $NN\to
NN\pi$, but some loops are to be promoted to significantly lower orders
compared to what is expected from Weinberg's original counting.  Ref.
\cite{norbertandme} left two questions unanswered that we will address in this
note:
\begin{itemize}
\item For the reaction channel $pp\to pp\pi^0$ the sum of all leading loops
  canceled; the origin of this cancellation could not be identified;
\item For the channel $pp\to d\pi^+$ the sum of diagrams of Fig.
  \ref{diagram}(b)--(d) gave a finite answer. However, as pointed out recently
  in Ref. \cite{anderstalk}, the corresponding amplitudes grow linearly with
  increasing final $NN$--relative momentum. This behavior leads to a large
  sensitivity to the final $NN$ wavefunction, once the convolution of those
  with the transition operators is evaluated as demanded by the
  non--perturbative nature of the $NN$ interaction.  The solution to this
  problem proposed in Ref. \cite{anderstalk} is to include a new counter term
  at leading order to absorb this unphysical behavior. However, chiral
  symmetry does not allow for such a structure (see Appendix for details).
\end{itemize}
As we will show in this paper, the solution to both questions are related and
at the same time shed some light on the concept of reducibility in pion
reactions
on few--nucleon systems.

We further demonstrate that the net effect of the inclusion of the NLO loops, shown in
 Fig. \ref{diagram}, is  to enhance the leading
rescattering amplitude by a factor of 4/3, bringing its contribution to the
cross section for $pp\to d\pi^+$ close to the experimental value.   

\section{Power counting and the concept of reducibility}

Already the existence of nuclei shows that perturbation theory is insufficient
to properly describe two--nucleon systems: only an infinite sum of diagrams
can produce a pole in the $S$--matrix. To bring this observation in line with
power counting, Weinberg proposed to classify all possible diagrams according
to the concept of reducibility \cite{weinNN,vk,evgeniNN}: those diagrams that
have a two nucleon cut are called reducible. Those which do not are called
irreducible. The latter make up the potential that is to be constructed
according to the rules of ChPT. The former are then generated by solving the
Schr\"odinger equation, using the mentioned 
potential as kernel. This scheme
acknowledges that the two nucleon cut contributions are enhanced compared to
the irreducible parts.

It was also Weinberg who gave a recipe how to calculate processes on few
nucleon systems with external probes \cite{swein1}: here the transition
operators are to be calculated using ChPT. Then those
transition operators must be convoluted with the appropriate $NN$ wave
functions --- in full analogy to the so--called distorted wave Born
approximation traditionally used in phenomenological calculations~\cite{kur}.

Therefore it is necessary to disentangle those diagrams that are part of the
wave function from those that are part of the transition operator. In complete
analogy to $NN$ scattering described above, the former are called reducible
and the latter irreducible. Also here the distinction stems from whether or
not the diagram shows a two nucleon cut. Thus, in accordance to this rule,
 the one loop diagrams shown in
Fig. \ref{diagram}(b)--(d) are irreducible, whereas diagrams (a)  seem to be
reducible. However, it will be the central finding of this work that diagrams
(a) contain a genuine irreducible piece due to the energy dependence of the
leading $\bar N N\pi\pi$--vertex, the so--called Weinberg--Tomozawa term (WT).
Specifically, the energy--dependent part of the WT vertex cancels one of the 
intermediate nucleon propagators, resulting in the irreducible part of diagrams (a).

\begin{figure}[t]
\begin{center}
\epsfig{file=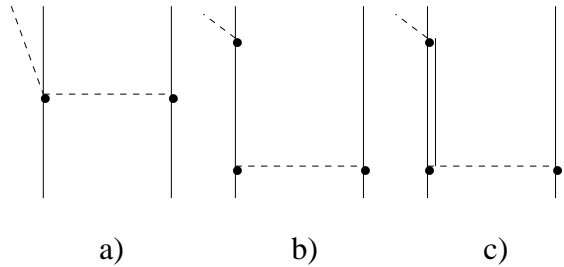, height=3.5cm, angle=0}
\caption{Tree level diagrams that contribute at leading ((a) and (b)) and
  next--to--leading order (c) to $NN\to NN\pi$. The double line denotes a
  $\Delta$--isobar. Note, in diagrams (b) and (c) --- for illustration --- with the
  one--pion exchange only one part
  of the $NN$ and $NN\to N\Delta$ potential is shown.}
\label{tree}
\end{center}
\end{figure}

As mentioned in the introduction, the power counting needs to be modified in
order to be applicable for $NN\to NN\pi$. The reason for this necessity is
the 
magnitude of the nucleon center-of-mass momentum $\vec p$ required to produce
a pion at rest in $NN$ collisions.  It is given by
\begin{equation} |\vec p\,| = \sqrt{m_\pi(M+m_\pi/4)}\,,\label{pin}\end{equation}
with $M=939\,$MeV and $m_\pi=139.6\,$MeV denoting the nucleon and pion mass,
respectively. Eq. (\ref{pin}) exhibits the important feature of the reaction
$NN\to NN\pi$, namely the large momentum mismatch between the initial and the
final nucleon-nucleon state. This leads to a large invariant (squared)
momentum transfer $t=-Mm_\pi$ between in- and outgoing nucleons. The
appearance of the large momentum scale $\sqrt{Mm_\pi}$ in pion production
demands a change in the chiral power counting rules, as pointed out already in
Ref.~\cite{bira1,pionprod}.  In addition, it seems compulsory to include the
Delta--isobar as an explicit degree of freedom, since the delta-nucleon mass
difference $\Delta = 293\,$MeV is comparable to the external momentum $p\simeq
\sqrt{M m_\pi} =362\, $MeV. The hierarchy of scales
\begin{equation}
m_\pi \ll p \simeq \Delta \ll M  \ ,
\label{schi}
\end{equation}
 suggested by this feature is
in line with findings within meson exchange models where the Delta--isobar gives
significant contributions even close to the threshold
\cite{jounidel,ourdelta}\footnote{For the channel $pp\to pp\pi^0$ strong
  support for an important role played by the $\Delta$--isobar was given by a
  partial wave analysis~\cite{deepak}.}.
The natural expansion parameter therefore is 
\begin{equation}
\chi = \frac{p}{M} = \sqrt{\frac{m_\pi}{M}} \ .
\label{expara}
\end{equation}
As a result at leading order only tree level diagrams contribute to the
transition operator (diagram (a) and (b) of Fig. \ref{tree}). Already at
next--to--leading order --- in addition to the first diagram that involves a
$\Delta$--isobar  (diagram (c) of Fig. \ref{tree}) --- the first loops appear (see
Fig. \ref{diagram}). As a consequence of the two scales $p$ and $m_\pi$ given in
Eq. (\ref{schi}) there exists a dimensionless parameter that is of
order $\chi$, namely $m_\pi/p$, that can appear as the argument of non--analytic 
functions as a result of the evaluation of loop integrals.
Thus, each loop now  contributes not only to a single order,
but to all orders higher than the one where it starts to contribute \cite{report}.
In this work we only consider the leading parts of the loops in Fig. \ref{diagram} that start 
to contribute at NLO.



%

At threshold only two amplitudes are allowed to contribute to the reaction
$NN\to NN\pi$, namely $A_{11}$
and $A_{10}$, where we used the notation $A_{T_iT_f}$ to label the total
isospin of the initial ($T_i$) and final ($T_f$) $NN$--pair. The third
amplitude allowed by the standard selections rules --- $A_{01}$ --- has to have at
least one $p$--wave in one of the final subsystems.  To the reactions $pp\to
pp\pi^0$ and $pn\to pp\pi^-$ only $A_{11}$  and  to $NN\to d\pi$ only
$A_{10}$ contribute at threshold, whereas
both $A_{11}$ and $A_{10}$ contribute to the reaction $pp\to pn\pi^+$.

The only transitions that are allowed to contribute near threshold are
$^3P_0\to {^1}S_0 s$ for $A_{11}$ and $^3P_1\to {^3}S_1s$ for $A_{10}$, where small letters denote the pion
angular momentum with respect to the $NN$ system and the $NN$ partial waves
are labeled with the standard notation $^{2S+1}L_J$. Those lead to the
following amplitude structures \cite{report}
\begin{eqnarray}
{\cal M}=iA_{11}\left((\vec S \cdot \vec p) \,{\cal I '}\right)
 +A_{10} \left((\vec S \times \vec p) \cdot \, \vec S \, '\right)
\label{structure}
\end{eqnarray}
with ${\cal I \,
  '}=\left(\chi_4^\dagger \sigma_y (\chi_3^T)^\dagger\right)/\sqrt{2}$ and
$\vec S = (\chi^T_1\sigma_y\vec \sigma\chi_2)/\sqrt{2}$ and $\vec S \, ' =
(\chi^\dagger_3\vec \sigma \sigma_y (\chi^\dagger_4)^T)/\sqrt{2}$, where the
$\chi_{1,2}$ ($\chi_{3,4}$) denote spinors for the incoming (outgoing)
nucleons.  For a deuteron in the final state we need to use
\begin{eqnarray}
{\cal M}= \tilde A_{10}\left((\vec S \times \vec p) \cdot
 \, \vec \epsilon_d\, ^* \right) \ ,
\label{structure2}
\end{eqnarray}
where now $\vec \epsilon_d \, ^*$ denotes the deuteron polarization vector
 and $\tilde A_{10}$ the convolution of
the production operator with the deuteron wave function---although the
 transition operators for $A_{10}$ and $\tilde A_{10}$ are the same, the
 matrix elements in general have different units. We come back to the
 evaluation
of the deuteron cross section
in section \ref{secresults}.
The amplitude in Eq. (\ref{structure}) is
normalized such that  for the total cross
section for $pp\to pn\pi^+$ we have
\begin{equation}
\sigma_{tot}(pp\to pn\pi^+) = \frac{M^3p}{16\pi^3\sqrt{s}}
\int \frac{dq'q'\, ^2p'}{\omega_{q'}}\left( \left|A_{11}\right|^2+
2\left|A_{10}\right|^2\right) \ ,
\end{equation}
where $q'$ denotes the cms momentum of the outgoing $\pi$, $p'$ the relative
momentum of the outgoing $pn$ pair, $s$ the invariant energy of the system,
and $\omega_q = \sqrt{q^2+m_\pi^2}$. Analogously, one finds for the 
 total cross
section for $pp\to d\pi^+$ 
\begin{equation}
\sigma_{tot}(pp\to d\pi^+) = \frac{2M_dM^2}{\pi s }q'p
\left|\tilde A_{10}\right|^2 \ ,
\end{equation}
where $M_d$ denotes the deuteron mass.

If we neglect all $NN$ distortions, we get for the leading rescattering
contribution (Fig. \ref{tree}(a)) at threshold \cite{norbertandme}
\begin{equation}
A_{11}^{2a} = 0 \, , \quad A_{10}^{2a} = 2\left(\frac{2m_\pi}{4f_\pi^2}\right)
\frac{1}{p^2-m_\pi^2}\left(\frac{g_A}{2f_\pi}\right)
=-\frac{g_Am_\pi}{2\vec p\, ^2f_\pi^3}(1+{\cal O}(\chi)) \ ,
\label{WT}
\end{equation} 
where we used that the $\pi N\to \pi N$ amplitude to leading order
contains  not only the standard WT term that scales as the sum of the incoming and
outgoing pion energies, here equal to $m_\pi$ and $m_\pi/2$, respectively, but
also its recoil correction equal to $\vec p\, ^2/2M = m_\pi/2$.  
The relevant terms of the underlying Lagrangian density are given in
the Appendix. Note that with a value of $2m_\pi/4f_\pi^2$ the WT  vertex including
the recoil correction as it appears in the $NN\to NN\pi$ amplitude takes
exactly the value it has for elastic $\pi N$ scattering at threshold.

\section{Evaluation of loops}

In the reaction $NN\to NN\pi$ the energies of the initial nucleons are of
order $\vec p\, ^2/2M\sim m_\pi$ and the momenta are of order $p$.  In
irreducible loops, on the other hand, both energies and momenta are of order
$p$ (see Appendix E of Ref. \cite{report} for details).  As a consequence in
the evaluation of diagrams (b),(c), and (d) of Fig. \ref{diagram} the nucleon
recoil terms (i.e. the nucleon kinetic energies) can be neglected in the
nucleon propagators, since they scale as $m_\pi$. At first glance this seems
to be at variance with the recent finding that three body $NN\pi$ cuts that
originate from the nucleon recoils play an essential role in pion reactions on
few nucleon systems \cite{recoils1,recoils2}. However, the reactions studied
in these references had very different kinematics, as they were small momentum
transfer reactions, where in typical kinematics the $\pi NN$ state was near
on--shell. Here, on the other hand, we are faced with a large momentum
transfer reaction: for typical kinematics a $\pi NN$ intermediate state is far
off--shell. In addition, here there is an additional kinematical suppression
for the $\pi NN$ cuts: at the cut the typical pion momentum, which set the
scale for the typical loop momentum, is that of the external pion of order of
at most $m_\pi$. As a consequence the $\pi NN$ cuts do not contribute before
N$^5$LO to the reaction $NN\to NN\pi$ \cite{report}.

  By comparison to the full results of
Ref.~\cite{dmitrasinovic}, in Ref.~\cite{norbertandme} it was shown that it is
allowed to expand the integrand of the loop integrals before evaluation in
powers of $\sqrt{m_\pi/M}$. As a consequence it was possible to express the
leading contribution of all loops corresponding to diagrams (b)--(d) of Fig.
\ref{diagram} in terms of a single integral
\begin{equation}
I_0(p_0,\vec p\, ^2)=\frac1i\int \frac{d^4l}{(2\pi)^4}\frac{1}
{(l_0-i\epsilon)(l^2-m_\pi^2)((l+p)^2-m_\pi^2)} \ .
\end{equation}
One finds $I_0(p_0,\vec p\, ^2)=I_0(\vec p\, ^2)
\left(1+{\cal O}\left(\chi\right)\right)=1/(16\sqrt{\vec p \, ^2})
\left(1+{\cal O}\left(\chi\right)\right)$.
The assumption of threshold kinematics (all outgoing momenta vanish)
 simplifies the operator structure significantly
and we can write --- neglecting for the moment the distortions from the $NN$
 interaction --- in order \cite{norbertandme}
\begin{eqnarray} \nonumber
A_{10}^{1b+1c+1d} &=& \frac{g_A^3}{16f_\pi^5} \left(-2+3+0\right)\,
\vec p \, ^2I_0(\vec p\, ^2)=\frac{g_A^3|\vec p|}{256f_\pi^5}  \\
A_{11}^{1b+1c+1d} &=& \frac{g_A^3}{16f_\pi^5}\left(-2+3-1\right)\, 
\vec p \, ^2I_0(\vec p\, ^2)=0 \ .
\label{results}
\end{eqnarray}
Note, here and in what follows we write equalities although we dropped terms
of higher order in $\chi$. As mentioned in the introduction, the sum of the
NLO loops vanishes in case of $A_{11}$. We will give an explanation for this
cancellation below. Let us now concentrate on $A_{10}$.  In order to compare
the result of Eq. (\ref{results}) to data, the transition operators need to be
convoluted with appropriate $NN$ wavefunctions. The convolution integrals that
arise necessarily involve non--vanishing $\vec p\, '$, denoting the outgoing
$NN$ relative momenta, even if we still work at threshold.  However, the
structure of the loop integral $I_0$ is such that --- to leading
order --- only $\vec p-\vec p\, '$ appears in the integrals and thus one can
directly generalize the expressions of Eqs. (\ref{results}).  As was argued in
Ref.~\cite{anderstalk}, this implies that for large $\vec p\, '$ the
contributions from the loops grows linearly with $|\vec p\, '|$. When it then
comes to the convolution of those operators with $NN$ wave function this
linear growth of the transition operators leads to a large sensitivity to the
deuteron wave functions.  However, there should be no sensitivity to the
particular wave functions used, for off--shell quantities are not observable
\cite{haag,lsz}.  On the other hand, the chiral Lagrangian does not allow for
a counter term to compensate this linear growth --- see Appendix for details.
The solution given in Ref. \cite{anderstalk}, namely the inclusion of a
counter term at leading order, is therefore not consistent with the
effective field theory used.  However, as we will show, the loops with the
unwanted behavior will be canceled exactly by an irreducible piece of diagrams
(a) of Fig.  \ref{diagram}. We proceed as follows: we first show this
cancellation to one loop order without distortions. Then we generalize the
result to the inclusion of the full $NN$ wave functions.

We still assume threshold kinematics and now turn to the evaluation of
diagrams (a) of Fig.  \ref{diagram}. 
In doing so one first has to realize that
in contrast to the irreducible diagrams discussed in the beginning of
this section, energies in the diagrams
with a two--nucleon cut are of the order of the external
energies ($l_0 \sim p^2/2M \sim m_{\pi}$).
Therefore, there is a priori no reason to neglect the nucleon recoils that are
of the order of $m_\pi$.  We thus get for the full expression for the first
diagram of Fig. \ref{diagram}(a), up to higher orders,
\begin{eqnarray} \nonumber
A^{1a1}_{10}\!\!\!\!&{=}&\!\!\!\!i\frac{3g_A^3}{32f_\pi^5}
\int \frac{d^4l}{(2\pi)^4}
\frac{[l_0+m_\pi-(2\vec p+\vec l)\cdot \vec l/(2M)]}
{(l_0-\frac{m_\pi}{2}-\frac{(\vec l+\vec
  p)^2}{2M}+i\epsilon)(-l_0+\frac{m_\pi}{2}-\frac{(\vec l+\vec
  p)^2}{2M}+i\epsilon)}
 \\ \nonumber
& & \qquad  \qquad \qquad \qquad \qquad \qquad \qquad \qquad \qquad \qquad \times
\frac{(\vec l\cdot(\vec l+\vec p))}{
(l^2-m_\pi^2)((l+p)^2-m_\pi^2)} \ , \\
A^{1a1}_{11}\!\!\!\!&{=}&\!\!\!\!0 \ ,
\end{eqnarray}
where we included the recoil correction to both the WT term in the numerator
as well as the nucleon energies in the denominator, in line with the
discussion above.  The vanishing $A^{1a1}_{11}$ reproduces the well known
result that the WT interaction does not contribute to the leading rescattering
diagram in $pp\to pp\pi^0$.

In order to proceed we rewrite the first term in the numerator of the above
integral as
$$
\left[l_0+m_\pi-\frac{(2\vec p+\vec l)\cdot \vec l}{2M}\right]=
\left[\left(l_0-\frac{m_\pi}{2}-\frac{(\vec p+\vec l)^2}{2M}\right)
+2m_\pi\right] \ ,
$$
where we used that at threshold $p^2/M=m_\pi$. The first term now exactly
cancels the first nucleon propagator and we are left with an expression that
no longer has a two nucleon cut --- it is irreducible. In this irreducible piece
we can neglect the recoil corrections in the remaining nucleon
propagator --- c.f. the discussion at the beginning of this section --- and get
\begin{eqnarray} \nonumber
A^{1a1}_{10}\!\!\!\!&{=}&\!\!\!\!i\frac{3g_A^3}{32f_\pi^5}
\int \frac{d^4l}{(2\pi)^4}\left\{
\frac{(\vec l\cdot(\vec l+\vec p))}{
(-l_0+i\epsilon)(l^2-m_\pi^2)((l+p)^2-m_\pi^2)}\right. \  \\
&+&\!\!\!\!
\frac{2m_\pi}{(l_0-\frac{m_\pi}{2}-\frac{(\vec l+\vec
  p)^2}{2M}+i\epsilon)(-l_0+\frac{m_\pi}{2}-\frac{(\vec l+\vec
  p)^2}{2M}+i\epsilon)}
\left.\frac{(\vec l\cdot(\vec l+\vec p))}{
(l^2-m_\pi^2)((l+p)^2-m_\pi^2)}\right\} \ . 
\end{eqnarray}
Up to higher orders the first term gives
\begin{equation}
A^{1a1 (\mbox{irr})}_{10}=-\left(\frac{3}{4}\right)
\frac{g_A^3}{16f_\pi^5}\vec p\, ^2I_0(\vec p\, ^2)
=-\frac{3}{4}\frac{g_A^3|\vec p|}{256f_\pi^5} \ ,
\label{irr1a1}
\end{equation}
where the label $(\mbox{irr})$ indicates that this is only the irreducible piece of
the diagram.
Analogous considerations for the second diagram of diagrams (a) of
Fig. \ref{diagram} give
\begin{eqnarray} \nonumber
A^{1a2}_{10}\!\!\!\!&{=}&\!\!\!\!i\frac{g_A^3}{32f_\pi^5}\int \frac{d^4l}{(2\pi)^4}\left\{
\frac{(\vec l\cdot(\vec l+\vec p))}{
(-l_0+i\epsilon)(l^2-m_\pi^2)(( l+ p)^2-m_\pi^2)}\right. \  \\
&-&\!\!\!\!
\frac{2m_\pi}{(l_0+\frac{m_\pi}{2}-\frac{(\vec l+\vec
  p)^2}{2M}+i\epsilon)(-l_0+\frac{m_\pi}{2}-\frac{(\vec l+\vec
  p)^2}{2M}+i\epsilon)}
\left.\frac{(\vec l\cdot(\vec l+\vec p))}{
(l^2-m_\pi^2)((l+p)^2-m_\pi^2)}\right\} \ .
\end{eqnarray}
The leading piece of the first term gives
\begin{equation}
A^{1a2 (\mbox{irr})}_{10}=-\left(\frac{1}{4}\right)
\frac{g_A^3}{16f_\pi^5}\vec p\, ^2I_0(\vec p\, ^2)
=-\frac{1}{4}\frac{g_A^3|\vec p|}{256f_\pi^5} \ .
\label{irr1a2}
\end{equation}
Thus we get 
\begin{eqnarray} 
\nonumber
A_{10}^{1a1(\mbox{irr})+1a2(\mbox{irr})+1b+1c+1d} &=& 
\frac{g_A^3}{16f_\pi^5}\left(-\frac{3}{4}-\frac{1}{4}-2+3+0\right)\,
\vec p \, ^2I_0(\vec p\, ^2)=0  \\
A_{11}^{1a1(\mbox{irr})+1a2(\mbox{irr})+1b+1c+1d} &=& 
\frac{g_A^3}{16f_\pi^5}\left(\phantom{-}0\,\,+0\,\,-2+3-1\right)\,
\vec p \, ^2I_0(\vec p\, ^2)=0 \ ,
\label{cancel}
\end{eqnarray}
where we repeat the results for $A_{11}$ from above for comparison. Thus, in
both channels that contribute at the production threshold the sum of all
irreducible loops that appear at NLO cancels.  On the other hand the remaining
pieces in the expressions for $A_{10}^{1a}$ exactly agree to the convolution
of the leading rescattering contribution with the one pion exchange, however,
with the $\bar N N\pi \pi$ WT vertex put on--shell. Thus the WT vertex takes
the value $2m_\pi/(4f_\pi^2)$ --- c.f. Eq. (\ref{WT}).  The two--nucleon
propagators in these integrals have a unitarity cut and it is this cut
contribution that should dominate the integral --- in line with Weinberg's
original classification as reducible and irreducible. In other words, these
pieces are indeed dominated by the reducible piece\footnote{In addition, the
  relative strength as well as sign in these terms equal to $-3$ for the pion
  exchange in the final $T_f=0$ channel compared to +1 for the pion exchange
  in the initial $T_i=1$ channel equal to the expectation values of the
  isospin parts of the one pion exchange in the relevant isospin channel,
  since $\left< TT_3|\vec \tau_1\cdot \vec \tau_2|TT_3\right>=2T(T+1)-3$.}.

\begin{figure}[t]
\begin{center}
\epsfig{file=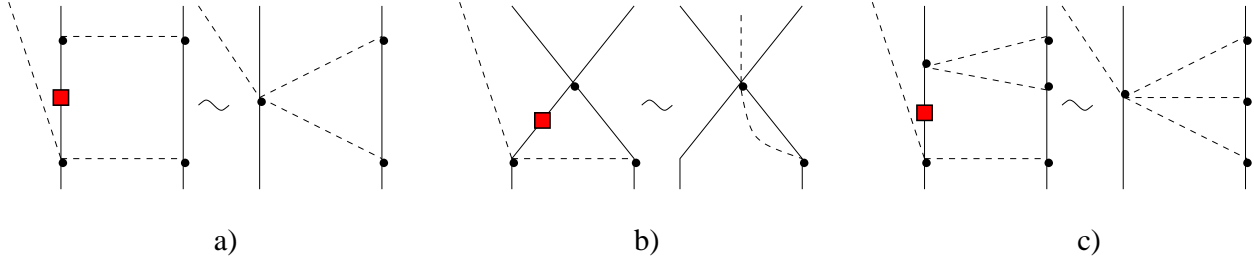, height=3.5cm, angle=0}
\caption{Illustration of the kind of topologies that corresponds to the
  irreducible structures (denoted by the filled box on the propagator that
  gets canceled by the energy dependence of the $\pi N\to \pi N$ vertex) that
  emerge from the convolution of the energy dependent rescattering term with
  various contributions to the $NN$ potential.}
\label{otherred}
\end{center}
\end{figure}

Next we show that for all ingredients of the $NN$ potential but the one--pion
exchange the choice of an on--shell $\bar N N\pi\pi$ vertex is of sufficient
accuracy. To this end we remind the reader that the integral corresponding to
the irreducible pieces of this first diagram of Figure \ref{diagram}(a) had a
structure like diagram (b) of that figure. This is illustrated in part (a) of
Fig.  \ref{otherred}. At leading order in the $NN$ potential there is besides
the one pion exchange also a contact interaction.  The convolution of the
rescattering diagram with the WT vertex (diagram (a) in Fig. \ref{tree}) with
this part of the $NN$ potential we can again decompose into a reducible piece
with the $\pi N\to \pi N$ vertex on--shell and an irreducible piece that takes
a structure of an integral with one nucleon less (see part (b) of Fig.
\ref{otherred}); this diagram, however, does not contribute below N$^4$LO and
is therefore irrelevant to the order we are working.  Thus, all that is to be
kept is the convolution of the on--shell rescattering contribution with the
contact $NN$ interaction --- in line with the findings of Ref.~\cite{toy}.  At
NLO in the $NN$ potential there are pion loops. Then the irreducible piece of
the convolution of the leading rescattering contribution with this piece
results in a two loop diagram (for a particular example see part (c) of Fig.
\ref{otherred}) that does not contribute up to N$^3$LO.  Thus, to the order we
are working we can safely put the WT vertex on--shell for the convolution of
any piece of the $NN$ potential with the leading rescattering contribution.

\begin{figure}[t]
\begin{center}
\epsfig{file=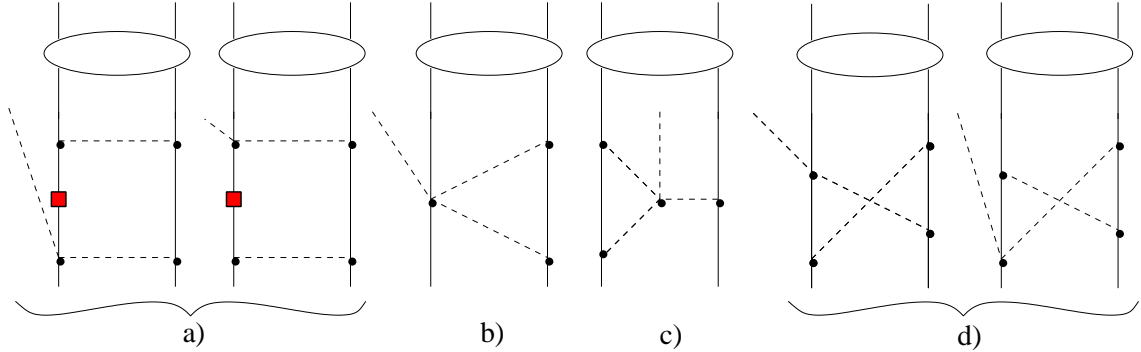, height=4.7cm, angle=0}
\caption{Leading loop diagrams for $NN\to NN\pi$, convoluted with the $NN$
  $T$--matrix, denoted by the ellipse. Here dashed lines denote
  pions and solid lines denote nucleons. The filled box on the nucleon propagators
  of the diagrams (a) indicate that only the irreducible piece is to be
  taken. The reducible part gets absorbed into the wave function.}
\label{diagram2}
\end{center}
\end{figure}

What remains to be shown is that the cancellation of Eq. (\ref{cancel})
survives (to the given order) the convolution with the full wave functions.
This generalization is straightforward. The corresponding diagrams are shown
in Fig. \ref{diagram2} for the inclusion of the final state interaction that
we want to discuss in detail. The argument in case of the initial state
interaction is completely analogous and will not be given.  Note that only the
irreducible parts of the diagrams (a) are to be included --- the reducible
pieces get absorbed into the wave functions.  Let $k$ denote the integration
variable of the convolution integral that we chose equal to the four momentum
of one of the nucleons. As argued in the previous paragraph, the integral will
be dominated by energies close to the corresponding on--shell energies. This
sets the scale for $k$--especially we can safely assume $k_0\ll p$. This is
all that is needed to neglect $k_0$ in the nucleon propagator of the pion loop
integrals. On the other hand in these loops $\vec k$ only enters as $\vec
p-\vec k$. Thus, the terms that enter in the convolution integrals with the
final state interaction to the order we are working are simply given by
replacing $\vec p$ by $\vec p-\vec k$ in Eqs.  (\ref{results}),
(\ref{irr1a1}), and (\ref{irr1a2}). This will obviously not change the
relative strength of the individual diagrams --- the cancellation survives the
convolution with the wave functions. We therefore conclude that up to
next--to--leading order all irreducible pion loops in the transition operator
cancel with the only effect  that 
the WT vertex in the rescattering diagram is to be put on--shell.

\section{Results}
\label{secresults}

Since the NLO diagrams discussed in this work contribute only to $A_{10}$, we
will now discuss their impact on the reaction $pp\to d\pi^+$ that is fully
determined by that amplitude. The cross section data for this reaction near
threshold is traditionally parameterized as
\begin{equation}
\sigma = \alpha \eta + \beta \eta^3 \, ,
\end{equation}
where Coulomb effects were neglected. To the reaction $pn\to d\pi^0$ both
$\alpha$ and $\beta$ contribute with only half their strength \cite{dpidata1}.
Here $\eta$ denotes the outgoing pion momentum in units of its mass.  The
first term gives the $s$--wave strength, whereas the second one denotes the
$p$--wave contribution (as well as some possible energy dependencies of the
$s$--wave~\cite{anderstalk}).

Before comparison with experiment is possible, the transition operators are
to be convoluted with appropriate $NN$ wave functions.  Here we use basically
the same formalism as described in Ref. \cite{recoils2}\footnote{For more details see
Appendix of Ref. \cite{TBK}.} and thus we do not
give any formulas in detail. For the $NN$ distortions we use the CD--Bonn
potential \cite{CDBonn} and use the same parameters as in Ref. \cite{recoils2}. To
calculate the leading order (LO) rescattering process (diagram (a) of Fig.
\ref{tree}) we use the standard expression for the WT term in threshold
kinematics --- thus we put $3/2 m_\pi$ at the vertex \cite{kur,chuck}. In
addition we also evaluate the direct contribution (diagram (b) of Fig.
\ref{tree}).  Since we derived the transition operator in threshold kinematics
we can only calculate $\alpha$. We obtain
\begin{equation}
\alpha^{LO} = 131 \ \mu\mbox{b} \ .
\end{equation}
This number is dominated by the rescattering contribution; switching off the
direct term alpha is lowered by 30 \% to $\alpha^{WT, \ LO}=101 \ \mu$b.
This value is consistent with those given in Ref. \cite{chuck}.
 Note that the direct term is known to
be quite model dependent \cite{chuck} and is small because of a 
cancellation of individually sizable terms. Clearly such a cancellation can
not be captured by the counting scheme. Still, we point out that in an EFT
scheme as used here all terms at a given order have to be retained.

As outlined above, in order to include the NLO contributions all we need to do
is to replace the $(3/2)m_\pi$ in the WT vertex by $2m_\pi$ --- 
or, stated
differently, to scale the given results for $\alpha^{WT, \ LO}$ by a
factor $(4/3)^2$. Thus, for the rescattering piece we get at NLO 
$\alpha^{WT, \ NLO}= 182 \ \mu$b, whereas the full result including
the direct term is
\begin{equation}
\alpha^{NLO}=220 \ \mu\mbox{b} \ .
\end{equation}
We checked that using different models for the $NN$ distortions changed the
rescattering contribution by about 10 \%, in line with previous findings~\cite{chuck}.

\begin{figure}[t]
\begin{center}
\epsfig{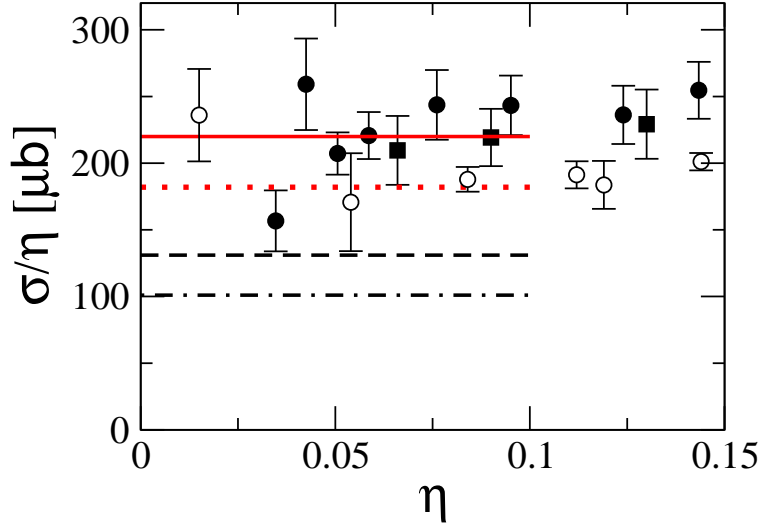}
\caption{Comparison of our results to experimental data for $pp\to d\pi^+$. 
The dashed and dashed--dotted
  curves show the LO results, where for the latter the direct contribution was
  omitted. The solid and dotted line show the results at NLO, where in the
  latter the direct term was omitted. The data is from Refs. \cite{dpidata1}
  (open circles), \cite{dpidata2} (filled circles) and \cite{dpidata3} (filled
  squares).}
\label{exp}
\end{center}
\end{figure}

In Fig. \ref{exp} we compare the results of our calculation to the
experimental data\footnote{Note that the data from $pn\to d\pi^0$
  \cite{dpidata1} are considerably lower than those for $pp\to d\pi^+$
  \cite{dpidata2,dpidata3}.  It appears unclear whether this discrepancy is due
  to systematic uncertainties (not shown in the figure) or due to other
  sources \cite{jounineu}.}. One clearly sees that going from LO to NLO
improves the description of the data. Note also that even a change by about a
factor of 2 in the cross section is in line with what is expected from the
counting: after all the expansion parameter for the amplitude is $\chi\sim
1/2$.

For a complete NLO calculation what needs to be considered in addition are
diagrams of the kind depicted in Fig. \ref{tree} (c), but with the full
$NN\to N\Delta$ transition amplitude. The evaluation of this diagram needs
as input the $NN\to N\Delta$ transition potential consistent with chiral
symmetry.
 The corresponding parameters are to be fixed from a fit to $NN$
data. Such a fit has not yet been performed and thus we postpone the
evaluation of diagram (c) to a later work. Note, however, that model
calculations show that the contribution of the Delta--isobar is not more than
10~\% in the amplitude \cite{ourdelta}.

\section{Summary and conclusions}

We have shown that the proper set of diagrams that contributes to the
transition operator for the reaction $NN\to NN\pi$ at NLO in chiral
perturbation theory is given by the diagrams of Fig. \ref{tree}, however, with
the $\bar N N\pi\pi$ vertex in diagram (a) put on--shell. To get to realistic
results these operators are to be convoluted with proper $NN$ wave functions.
The irreducible chiral loops that arise at this order exactly cancel those
terms that arise from the off--shell parts of the WT vertex. This cancellation
is required for formal consistency of the whole scheme, since the mentioned
diagrams show a linear growth with respect to the outgoing $NN$ momentum. Such
a growth would have led to a large sensitivity to the $NN$ wave function,
when the convolution with the final state interaction is calculated. This,
however, would have been in conflict with general arguments.

This at the same time also explains, why the sum of all loops has to vanish at
next--to--leading order for the reaction $pp\to pp\pi^0$: in this channel there
is no leading rescattering contribution. Thus, there is also nothing that could
cancel the linear divergence discussed above. The consistency
of the formalism therefore demands the sum of all loops to vanish.

As a result of our findings we can conjecture a general recipe on how to deal
with pion reactions in a nuclear environment in the presence of time
derivatives in vertices: one has to calculate all diagrams up to a given
order, including those that are formally reducible. Then the energy dependence
in the vertices is used to cancel one of the nucleon propagators. This
produces an irreducible piece that is to be part of the transition operator as
well as a reducible piece, where, however, the energy dependence of the
vertices is replaced by the corresponding on--shell value\footnote{Note that
  in the presence of quadratic time derivatives at individual vertices or the
  simultaneous appearance of several time derivatives in one diagram,
  additional vertices are to be included \cite{weind3}.}.

This new rule has significant impact on the role of isoscalar
rescattering in $NN\to NN\pi$ (diagram (a) of Fig. \ref{tree}, but with the
leading isoscalar interaction used for $\pi N\to\pi N$). Empirically the isoscalar $\pi N$ scattering
length is known to be very small. Theoretically it turned out that this
smallness is a consequence of an efficient cancellation amongst individually
large terms \cite{chiralpin}. Due to the energy dependence of the $\pi N\to
\pi N$ operators, however, when evaluated in the kinematics
relevant for pion production in $NN$ collision this cancellation is much less
efficient leading to a significant contribution from isoscalar rescattering
\cite{bira1,park}. If, on the other hand, the above rule is used, isoscalar
rescattering enters with the strength of the very small isoscalar $\pi N$
scattering length and thus would give a negligible contribution.

We have demonstrated that  the net effect of the inclusion of the NLO loops, shown in
 Fig. \ref{diagram}, is  to enhance the leading
rescattering amplitude by a factor of 4/3, bringing its contribution to the
cross section for $pp\to d\pi^+$ close to the experimental value.   

The next steps will be to evaluate the $NN\to NN\pi$ amplitudes to N$^2$LO for
both $s$-- and $p$--wave pions for all possible amplitudes.  At this order two
counterterms enter for the pion $s$--waves and one for the pion $p$--waves
both accompanied by $S$--wave nucleons in the final state. To this order
$p$--wave pions together with $P$--wave nucleons are parameter free
predictions.  On the other hand there are in total more than 40
observables\footnote{This large number is achieved by fully exploiting the 5
  dimensional phase space \cite{meyercomplete}.} measured for the reaction
channels $pp\to pp\pi^0$ \cite{meyercomplete}, $pp\to pn\pi^+$
\cite{daehnick}, $pp\to d\pi^+$ \cite{barbara} and $pn\to pp\pi^-$
\cite{heiko}.  Up to now only one phenomenological calculation was
compared to this large amount of data \cite{polres,report} and it was found
that all charged channels are well described, whereas there are significant
discrepancies for the neutral pions. It will therefore be of strong interest to
see if the new structures that emerge from the chiral Lagrangian are able to
cure these discrepancies.

Once the described channels are analyzed within ChPT one should move ahead to
consistently investigate the isospin violating observables measured recently,
namely the forward--backward asymmetry in $pn\to d\pi^0$ \cite{opper} as well as the total
cross section for $dd\to \alpha\pi^0$ \cite{ed}. First steps in this direction
were taken in Refs. \cite{jjb,ourddalpha}.

\vspace{1cm}

\noindent 
{\bf Acknowledgments}

\noindent 
We thank D.R. Phillips for a stimulating discussion.  We also thank the ECT*
in Trento, since this work was initiated during the workshop 'Charge Symmetry
Breaking and Other Isospin Violations' held there in June 2005.  This research
is part of the EU Integrated Infrastructure Initiative Hadron Physics Project
under contract number RII3-CT-2004-506078, and was supported also by the
DFG-RFBR grant no. 05-02-04012 (436 RUS 113/820/0-1(R)) and the
DFG--Transregionaler--Sonderforschungsbereich SFB/TR 16 "Subnuclear Structure
of Matter".  A.E.K, and V.B. acknowledge the support of the Federal Program of
the Russian Ministry of Industry, Science, and Technology No 40.052.1.1.1112.

\appendix

\section{Lagrange densities and vertices}

The starting point is an appropriate Lagrangian density, constructed to be
consistent with the symmetries of the underlying more fundamental theory (in
this case QCD) and ordered according to a particular counting scheme.
Omitting terms that do not contribute to the order we will be considering
here, we therefore have for the relevant terms of the leading and
next--to--leading order Lagrangian in sigma gauge for the pion field
\cite{vk}
\begin{eqnarray}\nonumber
 {\cal L}\!\!\!\! & = & \!\!\!\!
          \frac{1}{2}\partial_\mu{\boldpi}\partial^\mu{\boldpi}
          {-}\frac{1}{2}m_{\pi}^{2}\boldpi^{2}
          {+}\frac{1}{6f_\pi^2}\left[
(\boldpi \cdot \partial_\mu \boldpi)^2{-}\boldpi^2(\partial^\mu
\boldpi \cdot \partial_\mu \boldpi) \right]
 \\ 
    &   &  \!\!\!\! {+}N^{\dagger}[i\partial_{0}{-}\frac{1}{4 f_{\pi}^{2}} \boldtau \cdot
         (\boldpi\times\dot{\boldpi})]N{+}\frac{g_{A}}{2 f_{\pi}} 
         N^{\dagger}\boldtau\cdot\vec{\sigma}\cdot\left(\vec{\nabla}\boldpi
{+}\frac{1}{2f_\pi^2}\boldpi(\boldpi \cdot \vec \nabla \boldpi)
\right)N 
                                               \nonumber \\
& & \!\!\!\!{+}\frac{1}{2m_{N}}[N^{\dagger}\vec{\nabla}^{2}N
{+}\Psi_\Delta^{\dagger}\vec{\nabla}^{2}\Psi_\Delta]                            
        {+}\frac{1}{8M_N f_{\pi}^{2}}(iN^{\dagger}\boldtau\cdot
        (\boldpi\times\vec{\nabla}\boldpi)\cdot\vec{\nabla}N {+} h.c.)
                                       \nonumber \\ 
  &   &  \!\!\!\!{-}\frac{g_{A}}{4 m_{N} f_{\pi}}[iN^{\dagger}\boldtau\cdot\dot{\boldpi}
        \vec{\sigma}\cdot\vec{\nabla}N {+} h.c.]             
       {-}\frac{h_{A}}{
        2 m_{N} f_{\pi}}[
        iN^{\dagger}\boldT\cdot\dot{\boldpi}\vec{S}\cdot\vec{\nabla}
        \Psi_\Delta {+} h.c.]    \     
%
  \nonumber \\
  &   &  \!\!\!\!-\frac{d_1}{f_{\pi}} 
        N^{\dagger}(\boldtau\cdot\vec{\sigma}\cdot\vec{\nabla}\boldpi)N\,
        N^{\dagger}N     -\frac{d_2}{2 f_{\pi}} \varepsilon_{ijk} \varepsilon_{abc} 
        \partial_{i}\pi_{a}  
        N^{\dagger}\sigma_{j}\tau_{b}N\, N^{\dagger}\sigma_{k}\tau_{c}N 
        +\cdots \ ,              \label{la1}
\end{eqnarray}
\noindent
where $f_\pi$ denotes the pion decay constant in the chiral limit, 
$g_A$ is the 
axial--vector coupling of the nucleon.
 $h_A$ is the $\Delta N \pi$ coupling, and
$\vec S$ and $\boldT$ are the transition spin and isospin
matrices, normalized such that
\begin{eqnarray}
S_iS_j^\dagger = 
\frac{1}{3}(2\delta_{ij}-i\epsilon_{ijk}\sigma_k) \, , \ 
T_iT_j^\dagger =
 \frac{1}{3}(2\delta_{ij}-i\epsilon_{ijk}\tau_k) \ .
\end{eqnarray}
The dots symbolize that what is shown are only those terms that are relevant
for the calculations presented.  As demanded by the heavy baryon formalism,
the baryon fields $N$ and $\Psi_\Delta$ are the velocity--projected pieces of
the corresponding relativistic fields; e.g. $N=1/2(1 {+} \barr v )\psi$, where
$v^\mu=(1,0,0,0)$ denotes the nucleon 4--velocity.  The corresponding vertex
functions can be read off directly from Appendix A of Ref. \cite{ulfs}.  The
relevant vertices for the $NN$ interaction are discussed in Ref.
\cite{evgenirev}.

The last two terms in Eq. (\ref{la1}) show the leading four--nucleon--pion
counter terms.  The Pauli principle for the $NN$ system only allows them to
contribute in one fixed linear combination. The corresponding operator
contributes to pion $p$--waves in $pp \to
pn\pi^+$ as well as to the leading three--body force \cite{pionprod,evgeni3N}.
To $NN\to NN\pi$ at threshold this operator can only contribute through loops.
The leading four--$N$--$\pi$ vertices that contribute there are suppressed by
an additional $m_\pi/M$ \cite{bira1,park}.  In general, to be in line with the
Goldstone theorem, it is a necessary requirement that counter terms either
contain a derivative acting on the pions, or scale with $m_\pi^2$. The latter
is a consequence of the well known relation $m_\pi^2 \propto \hat m$, where
$\hat m$ denotes the average mass of the light quarks. To be consistent with
chiral symmetry the chiral Lagrangian is only allowed to have terms analytic
in $\hat m$. To absorb the sensitivity to the wave function discussed in the
introduction would require a counter term that neither contains powers of
$m_\pi$ nor a derivative acting on the pion field at variance with chiral
symmetry.

\pagebreak

\end{document}